# 11-μs Time-resolved, Continuous Dual-Comb Spectroscopy with Spectrally Filtered Mode-locked Frequency Combs

Nazanin Hoghooghi*, Ryan K. Cole, Gregory B. Rieker

[1]Precision Laser Diagnostics Laboratory, Department of Mechanical Engineering,

University of Colorado Boulder, Boulder, CO 80309, USA

*Corresponding author: nazanin.hoghooghi@colorado.edu

**Abstract:**

Broadband dual-comb spectroscopy (DCS) based on portable mode-locked fiber frequency combs is a powerful tool for in situ, calibration free, multi-species spectroscopy. While the acquisition of a single spectrum with mode-locked DCS typically takes microseconds to milliseconds, the applications of these spectrometers have generally been limited to systems and processes with time changes on the order of seconds or minutes due to the need to average many spectra to reach a high signal-to-noise ratio (SNR). Here, we demonstrate high-speed, continuous, fiber mode-locked laser DCS with down to 11 μs time resolution. We achieve this by filtering the comb spectra using portable Fabry-Perot cavities to generate filtered combs with 1 GHz tooth spacing. The 1 GHz spacing increases the DCS acquisition speed and SNR for a given optical bandwidth while retaining a sufficient spacing to resolve absorption features over a wide range of conditions. We measure spectra of methane inside a rapid compression machine throughout the 16 ms compression cycle with 133 $cm^{-1}$ bandwidth (4000 comb teeth) and 1.4 ms time resolution by spectrally filtering one of the combs. By filtering both combs, we measured a single-shot, 25 $cm^{-1}$ (750 comb teeth) spectrum of CO around 6330 $cm^{-1}$ in 11μs. The technique enables simultaneously high-speed and high-resolution DCS measurements, and can be applied anywhere within the octave-spanning spectrum of robust and portable fiber mode-locked frequency combs

## I. Introduction

Laser absorption spectroscopy (LAS) is a nonintrusive spectroscopic technique which has enabled quantitative, in situ measurements of temperature, pressure and species mole fraction in a variety of environments. Optical frequency combs are broadband laser sources with evenly spaced spectral elements ('comb teeth'). Frequency combs have been shown to be powerful sources for LAS, enabling simultaneous multi-spices detection as well as pressure and temperature measurements with high accuracy over a wide range of measurement conditions [1–12].

Dual comb spectroscopy (DCS) uses a multi-heterodyne detection scheme that enables fast, comb-tooth-resolved spectroscopy with a single photodetector. Two optical frequency combs with different repetition rates, $f_{rep}$ and $f_{rep}+ \Delta f_{rep}$, are interfered on a detector.

Absorption information encoded on individual comb teeth is down converted from the optical domain to the RF domain through the beating of pairs of comb teeth [13–15].

The achievable spectral point spacing, optical bandwidth, and acquisition time of a single spectrum depends on the comb repetition rates. The point spacing for a single spectrum is equal to the spacing between the comb teeth, which is set by the pulse repetition rate of the mode-locked combs. There are methods to scan the comb between the acquisition of individual spectra to 'fill in' and increase the resolution of the spectrum, but this requires acquisition of multiple spectra [16–18] and has so far not been shown at microsecond timescales.

The achievable optical bandwidth for a single spectrum is not set by the bandwidth of the combs, but rather the spectral window that is defined at one end by a location where the comb teeth of the two combs exactly overlap, and at the other end by a location where a tooth from one comb is equidistant from the nearest two teeth of the other comb. The light beyond this window must be optically filtered before the detector to avoid aliasing different parts of the spectrum during the RF downconversion. The bandwidth of this window, $\Delta\upsilon$, depends on the comb repetition rate and repetition rate difference as defined in Eq. 1 [15].

$$\Delta v \leq \frac{f_{rep}^2}{2\,\Delta f_{rep}} \qquad \text{Eq.1}$$

The acquisition time for a single, full-resolution spectrum, $\tau$, is equal to the inverse of the difference between the two comb repetition rates, as shown in Equation 2.

$$\tau = \frac{1}{\Delta f_{rep}} \qquad \text{Eq.2}$$

We can see that a higher repetition rate difference ($\Delta f_{rep}$) leads to faster acquisition of a full spectrum, but also decreases the achievable optical bandwidth for a given comb tooth spacing ($f_{rep}$). However looking at the combination of these equations, we can also see that it is possible to enable faster acquisition without sacrificing optical bandwidth by increasing both the comb repetition rate and the repetition rate difference, though at the expense of increasing the spectral point spacing (thereby decreasing the ability to resolve absorption features).

Thus, there are tradeoffs between bandwidth, point spacing and speed that must be adjusted to suit the requirements of the system being measured. In many sensing applications, the DCS approach needs to have broad bandwidth and a reasonably small comb tooth spacing in order to accurately measure mole fraction, temperature and pressure. This is especially important for sensing in dynamic environments (e.g. in combustion systems such as engines or shock tubes) that are subject to variable pressures and temperatures ranging from atmospheric conditions to very high pressures and temperatures, and with both large and small molecules present. In such systems, the optimal repetition rate balances a sufficiently high spectral resolution to resolve narrow absorption features near atmospheric conditions with a sufficient time resolution to capture the transient nature of the system. To clarify this point, we simulate the spectrum of methane between 5996-6010 $cm^{-1}$ at two different pressures (1 bar and 25 bar) and vary the spectrum point spacing. At 1 bar, decreasing the point spacing from 200 MHz to 30 GHz dramatically affects the retrieved spectrum, which contain features with linewidths on the order of 3-5 GHz (Fig 1 (a)). However, at 25 bar, decreasing the point spacing does not affect the already broadened lines (Fig (b)). In this example, if the system under study has a dynamic pressure range of 1-25 bar, a comb tooth spacing or spectral resolution of 1-2 GHz is sufficient to resolve the absorption features over the entire pressure range. A comb tooth spacing greater than 2 GHz will allow for faster acquisition but significantly undersamples the absorption features at low pressures. Thus in this example, and for many examples involving atmospheric pressure gases, a pulse repetition rate (comb tooth spacing) of 1-2 GHz is ideal and will maximize bandwidth and acquisition speed while still resolving complex absorption spectra.

Despite their appeal for high-speed sensing applications, there are currently no comb sources suitable for field measurements with 1-2 GHz repetition rates (discussed in more detail below). In this work, we generate 1 GHz broadband frequency combs by spectrally filtering two 200 MHz repetition rate fiber mode-locked lasers. We take advantage of existing robust and portable, lower repetition rate fiber mode-locked comb technology that routinely generates octave-spanning spectra. This optical bandwidth is more difficult to generate directly from sources with GHz repetition rate. Using spectral filters constructed from optical cavities containing mirrors with broadband coatings and low dispersion, and by using a flexible cavity-locking scheme, we are able to select (and change) the measurement window of interest to anywhere within the broad spectrum of the fiber frequency comb. Increasing the repetition rate of these lasers by using portable spectral filters allows for faster individual spectrum acquisition time (through higher $\Delta f_{rep}$) for the same optical bandwidth, and higher power per comb tooth before detector saturation for an increased single-spectrum SNR [19]. We demonstrate the application of the 1 GHz frequency combs for ms and μs DCS in a rapid compression machine and in gas cell, respectively.

The organization of the paper is as follows: Section II is a short overview of the existing comb sources for high-speed DCS. In Section III we discuss two mode filtering techniques for improving the practical time resolution of DCS. In Section IV, we demonstrate a high-speed DCS measurement with 1.4 ms time resolution in a rapid compression machine (RCM) using one spectrally filtered frequency comb. Finally, in Section V, we show that by increasing the repetition rate of both combs from 204 MHz to 1.02 GHz, the time resolution of high-SNR DCS is further improved to 11 μs.

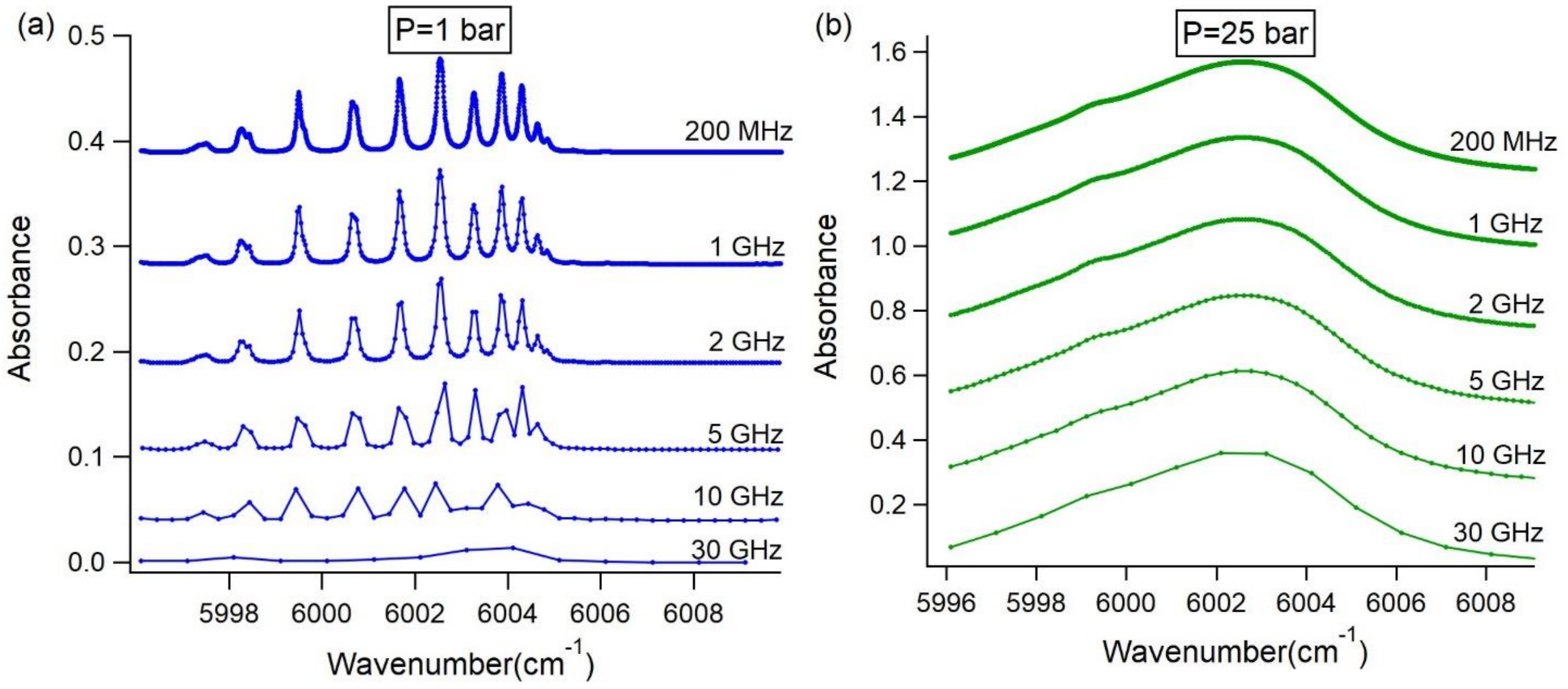

Fig.1. Modeled spectra of methane ($CH_4$) plotted with different spectral point spacing (offset for clarity). (a) Spectra modeled at 1 bar (b) Spectra simulated at 25 bar, 300K, with 5mm pathlength and 0.5 mole fraction.

## II. Existing Comb Sources for High-Speed DCS

The comb tooth spacing and power-per-tooth depend on the frequency comb source. There are three major classes of optical frequency comb sources that have been used for DCS: fiber mode-locked laser combs [3,6,8,10,20–28], EOM combs [29–31], and chip-based sources such as microresonators [32–34], quantum cascade lasers (QCL) [4,35] or intraband cascaded lasers (ICL) [36].

Fiber mode-locked frequency combs typically generate <100 fs pulses with 50-300 MHz repetition rate, which after amplification can generate a broad spectral bandwidth covering >3000 cm$^{-1}$ made up of hundreds of thousands of comb teeth. Due to their large number of finely spaced comb teeth, the power per comb mode is typically low (on the order of nW per comb mode) which results in a low single-spectrum SNR and typically imposes the need to average over seconds to minutes to achieve a high SNR spectrum. Although methods for timing repetitive events with the comb pulses can be used to interleave spectra and demonstrate high DCS time resolution (e.g. [3]), these techniques do not offer continuous measurements.

EOM combs are frequency agile comb sources since their comb tooth spacing can be adjusted by changing the electro-optic modulator driving frequency from a few MHz to 10s of GHz. The native optical bandwidth of these combs is typically only a few nm (a few hundred comb teeth) and generation of a broadband spectrum comparable to fiber mode-locked lasers is challenging [37,38]. The center wavelength of EOM combs is generally limited to the 1550 nm region due to the availability of low $V_\pi$ modulators in this band. Overall, EOM combs are capable of high SNR DCS measurements with few-ms acquisition time but over a narrow bandwidth that covers a few absorption features in the NIR [30,39].

Chip-based frequency comb sources have been the subject of intensive research. DCS with microresonator soliton combs has been demonstrated [32–34]. These combs have broad optical bandwidth and large comb tooth spacing on the order of 10s of GHz. The large comb tooth spacing allows for acquisition times on the order of nanoseconds or microseconds, however it limits their application for gas-phase spectroscopy. Other newly invented chip-based combs are QCL and ICL combs which directly emit light in the mid-IR region [36,40]. Fast DCS measurements with single-micro-second acquisition times has been demonstrated with QCLs. However QCL combs have very narrow bandwidth of up to 100 $cm^{-1}$ and large comb tooth spacing in the order of 10 GHz which limits the range of their application to mostly high pressure or broadly absorbing systems [4,41].

Despite their potential benefit in many high-speed sensing applications, broadband frequency combs with 1-2 GHz repetition rates are not yet readily available for field measurements. 1 GHz frequency combs based on Ti:Sapphire [42], Yb [43] and monolithic Er-glass [44] have been demonstrated, among which only Ti:Sapphire frequency combs are commercially available, generating light between 520 nm to 1200 nm [45]. Recently a 1 GHz mode locked laser at 1550 nm has been commercially offered that has the potential to be fully stabilized for comb operation in the C-band [46].

## III. Spectral filtering for high-speed DCS

In this paper, we generate a 1 GHz repetition rate frequency comb from a 200 MHz repetition rate Er-fiber frequency comb by spectral filtering with a Fabry-Perot cavity. If the ratio of the cavity free spectral range (FSR) to comb repetition rate (and therefore comb tooth spacing) is an integer $m$, then only every $m$th comb tooth passes through the cavity, and the comb spacing is increased by a factor of $m$. Spectral filtering can be applied to one or both frequency combs in a DCS system to enable high-speed measurements anywhere within the native Er-fiber comb spectrum, which can be easily broadened to cover 1-2 µm owing to the lower repetition rates. Figure 2 shows a typical DCS detection scheme (Fig 2(a)), and the schemes including one or two optical filters (Fig. 2 (b) and(c)).

As mentioned in Section I, increasing the repetition rate of the combs influences the DCS point spacing, optical bandwidth, or acquisition time. Both cavity-filtering schemes enable broadband measurements at faster single-spectrum acquisition times than can be achieved with unfiltered combs covering the same optical bandwidth. In this case, the single-spectrum acquisition time is shortened by factors of $m$ and $m^2$ for the one- and two-filter cases respectively (see Eq. 1).

In addition to faster single-spectrum acquisition times for a given bandwidth, increasing the comb repetition rates can also impact the overall acquisition time to reach a desired SNR. According to Eq. (2) in Ref [47], as we decrease the number of comb teeth by $m$ and while keeping the total comb power hitting the detector constant (i.e. increasing power-per-comb tooth), the single-spectrum SNR is improved by a factor of $m$. In addition, there is evidence that increasing the repetition rate of the lasers extends the linear operating range of the photodetector [19,48]. As a result, higher overall optical powers in comparison to the unfiltered case can be placed on the detector before saturation which further increases the SNR when operating in the detector noise or shot-noise dominated regime. Here, because both cavity filtering schemes increase the repetition rates and reduce the overall number of comb teeth, they each result in a higher single-spectrum SNR as a result of higher power-per-comb-tooth that can be placed on the detector before saturation. This results in the need for fewer averages of successive spectra to achieve a particular SNR. The improvement in minimum acquisition time is not the same for both filtering cases. In the one-filter case, the power-per-comb tooth of only the filtered comb can be increased. Conversely, in the two-filter case, each comb can have a higher power-per-comb tooth resulting in a higher single-spectrum SNR than the one-filter case.

Here, it is important to note that although filtering enables higher single-spectrum SNR, it could negatively affect the uncertainty of a quantitative fit between the measured spectrum and absorption model. Although fit uncertainty is nominally related to the square root of number of comb teeth in the spectrum, in reality it depends on how many comb teeth carry information by sampling absorption features. Due to this complex relation between fit uncertainty, number of comb teeth, and the conditions of the measurement (which dictates how many comb teeth overlap with absorption features), developing a simple analytical equation for the overall gain from filtering is not straightforward. As a result, the exact filtering ratio (FSR/$f_{rep}$) must be chosen not only based on the required spectral point spacing and SNR but also in consideration with required measurement uncertainty. For this reason, certain applications may need to strike a balance between the single-spectrum SNR and the total number of comb teeth in the spectrum to achieve the lowest overall uncertainty.

Experimentally, the one-filter approach is simpler to implement than the two-filter approach. The simplicity of implementation is due to the fact that no matter which set of comb teeth are filtered, there is always a set of comb teeth between the filtered and the unfiltered comb that will generate heterodyne beats between 0 and $f_{rep}$ /2. Put another way, any set of filtered comb teeth will satisfy the Nyquist sampling condition (Eq. 1) at the expected optical frequencies, the only difference being that the spectrum now consists of m-times fewer RF beat tones. This greatly relaxes the constraints on the cavity lock since no specific comb tooth set needs to be filtered, as will be shown in section IV.

When both frequency combs are filtered, even higher single-shot SNR is possible along with a faster acquisition time compared to the unfiltered case with the same optical bandwidth. However, the two-filter approach is more challenging to implement experimentally than the one-filter method since it requires control over which sets of comb teeth are filtered. This is necessary to maintain the correct mapping from the optical to the RF domain. If filtering is done randomly, it may result in a significant shift of the RF heterodyne beat notes or aliasing. In this case, the construction of the spectrum from the measured RF beat notes could be challenging or in some cases not possible. This added complexity means that a more complex cavity locking scheme needs to be implemented in order to have unambiguous control over which comb teeth are filtered, as will be shown in section V.

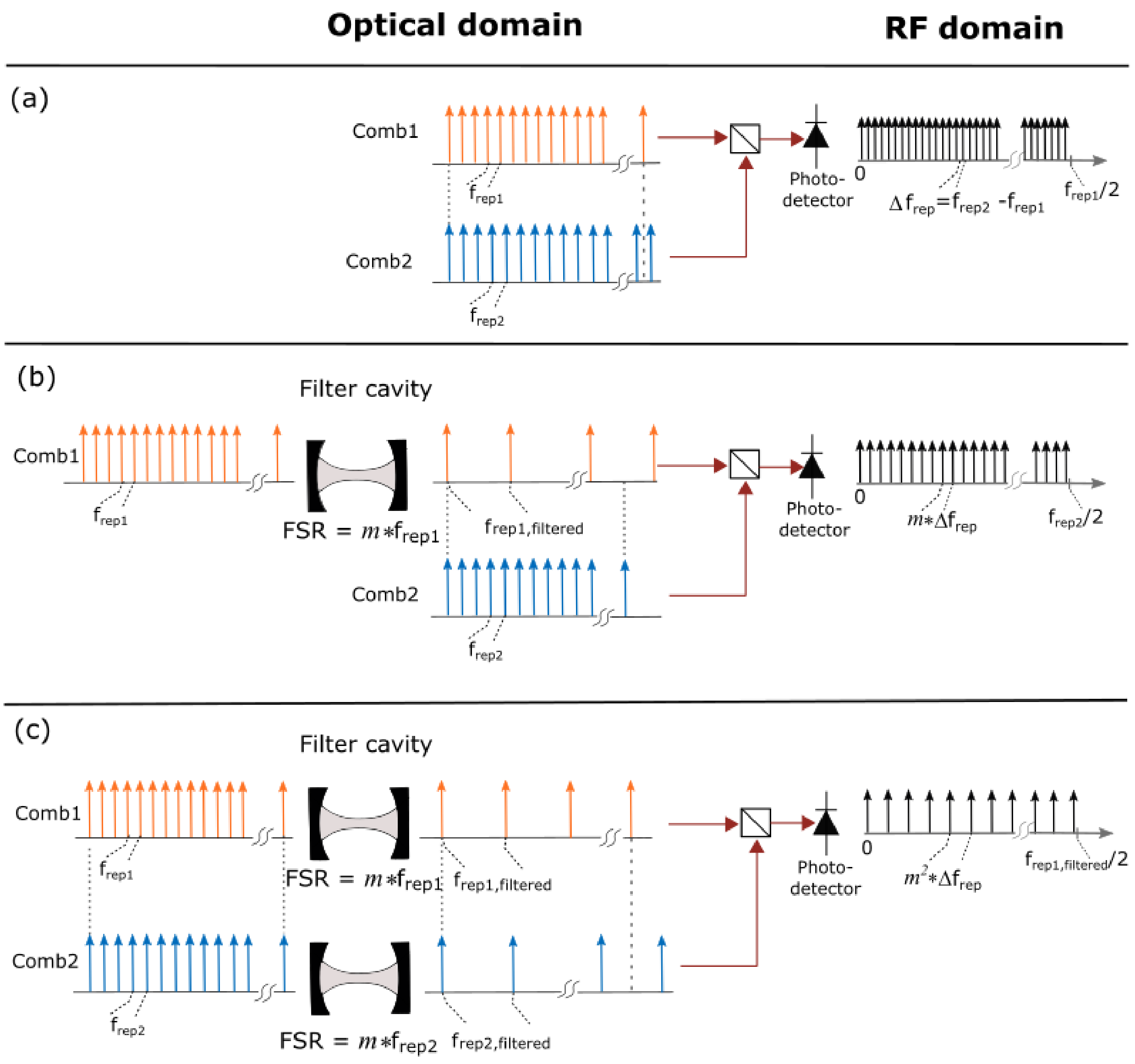

Figure 2. (a) Typical DCS scheme where the two combs repetition rates differ by $\Delta f_{rep}$. (b) DCS with one spectrally filtered comb. The optical bandwidth of the acquisition window is the same as without a filter, but the spacing between the heterodyne beats is $m*\Delta f_{rep}$, giving a factor of m increase in single-spectrum acquisition time. (c) DCS with two spectrally filtered combs. Again, the optical bandwidth of the acquisition window is the same as without a filter, but the spacing between heterodyne beats is $m^2*\Delta f_{rep}$ giving a factor of $m^2$ faster acquisition time than (a). The

## IV. High-speed DCS in a Rapid Compression Machine

To demonstrate the applicability of the spectral filtering method to real experimental conditions, we performed a proof-of-concept measurement inside a rapid-compression machine (RCM). RCMs are used to simulate a single compression stroke of an internal combustion engine. They provide an ideal testbed for validating LAS sensors in realistic, engine-like conditions. An RCM typically consists of one or two pneumatically driven pistons opposing a gas volume [49]. The release of the pistons compresses the gas sample, increasing the pressure and temperature over a period of a few ms. Broadband DCS using mode-locked lasers is an ideal sensor for measurement over the wide range of pressure (e.g. from 1 bar to 80 bar) and temperature (e.g. from 290 to 1000 K) of the RCM. However, the time resolution of DCS has to be improved to the millisecond level or better so that multiple spectra can be recorded over the short compression process.

To achieve the required millisecond time resolution, the measured DCS data can be apodized in post processing [5] to reduce measurement noise or the comb tooth spacing can be increased using mode filters, the approach taken here. We chose to filter only one of the frequency combs for two reasons. First, it is simpler and more robust to establish the lock for one filter cavity given the noisy environment. Second, the required time resolution for this experiment is milliseconds, which relaxes the single-shot SNR requirement and allows for coherent averaging of a few spectra. We used a mobile DCS system similar to the one described in [8,21]. The frequency combs are two polarization-maintaining fiber mode-locked lasers with center wavelength at 1550 nm and comb tooth spacing of ~204 MHz. The light from one of the frequency combs was sent to a filter cavity. The cavity is made of two curved mirrors (R=50 cm) with finesse of ~300. The cavity is ~14.7 cm long which corresponds to an FSR of 1.02 GHz. The cavity was locked to the input frequency comb using a dither locking scheme, which remained locked during the compression despite the loud noise and associated vibrations of the machine. The low cavity finesse and low group delay dispersion (GDD) of <± 5 $fs^2$ allow for almost the entire comb spectrum (~1450 $cm^{-1}$) to be spectrally filtered. The light from the filtered and unfiltered comb were combined using a 50/50 fiber coupler and sent to the RCM. The difference between the repetition rates of the combs after filtering is ~ 14 kHz, i.e each spectrum was collected in 71 μs.

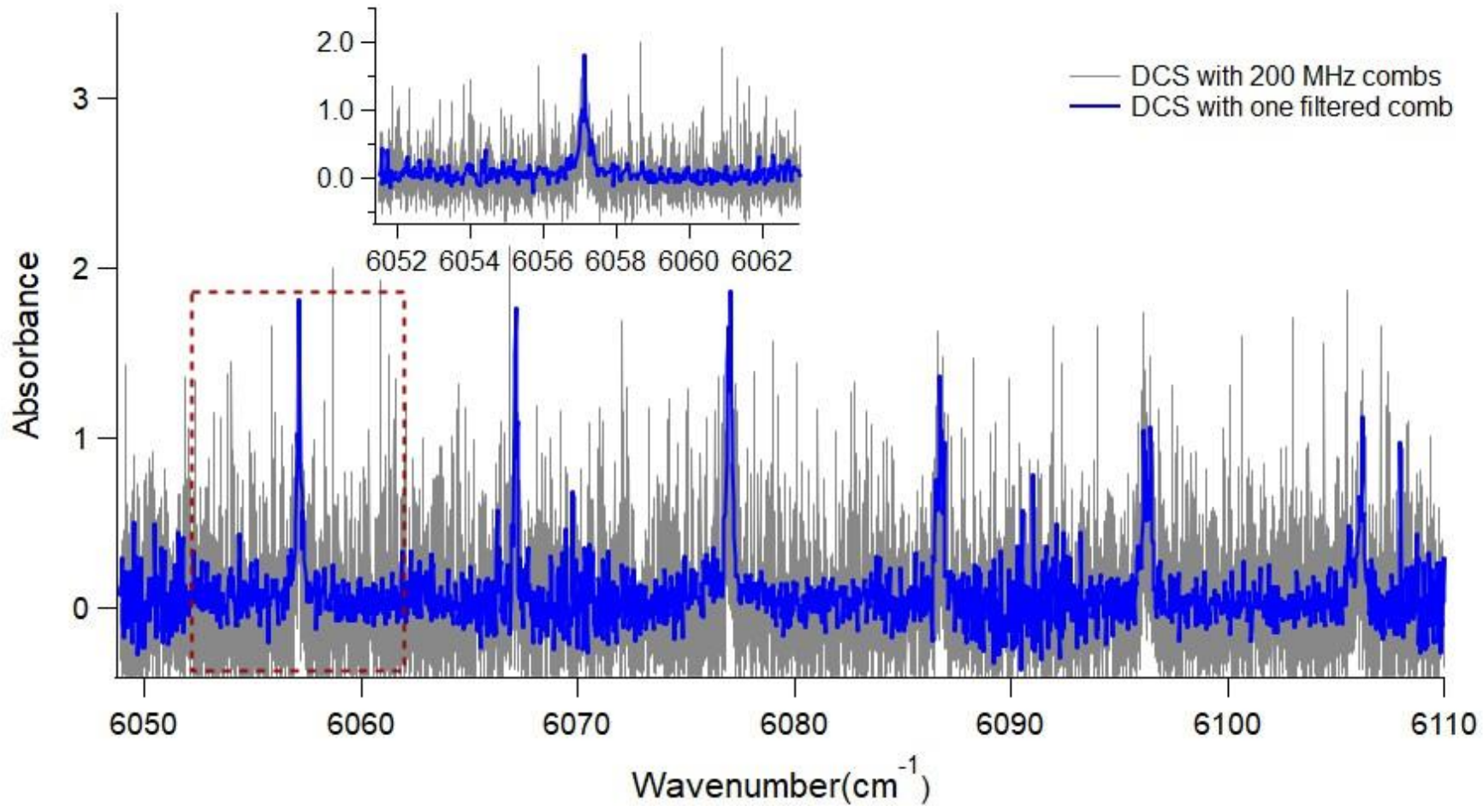

Fig.3 Methane spectrum measured with the one filtered comb DCS (blue) and with the 200 MHz DCS (gray) averaged for 1.4 ms (20 interferograms).

The combustion chamber was 4.63 cm long and filled with a gas mixture of $CH_4$ (75%) and dry air. After passing through the chamber, the light was focused on a fast photodetector (150 MHz bandwidth) and the signal was digitized using a 250 MS/s data acquisition card. Figure 3 shows the baseline corrected absorbance spectrum of methane measured using this technique in 1.4 ms (blue trace). To showcase the improvement in SNR by using this technique, we overlaid the absorbance signal obtained without mode filtering (gray trace). The SNR of the DCS signal with one filtered comb is ~3.7 times higher than the unfiltered case.

The 1.4-ms DCS time resolution obtained by spectrally filtering one of the combs allows us to measure the spectrum of $CH_4$ over the entire compression cycle of ~16 ms at 1.4 ms intervals. Figure 4 shows example recorded absorbance spectra during the compression every 1.4 ms from the start of the compression. The broadening of the lines due to the increased pressure can be clearly seen. These results show the improvement of the DCS time resolution from seconds to ms by using a simple and portable mode-filtering technique.

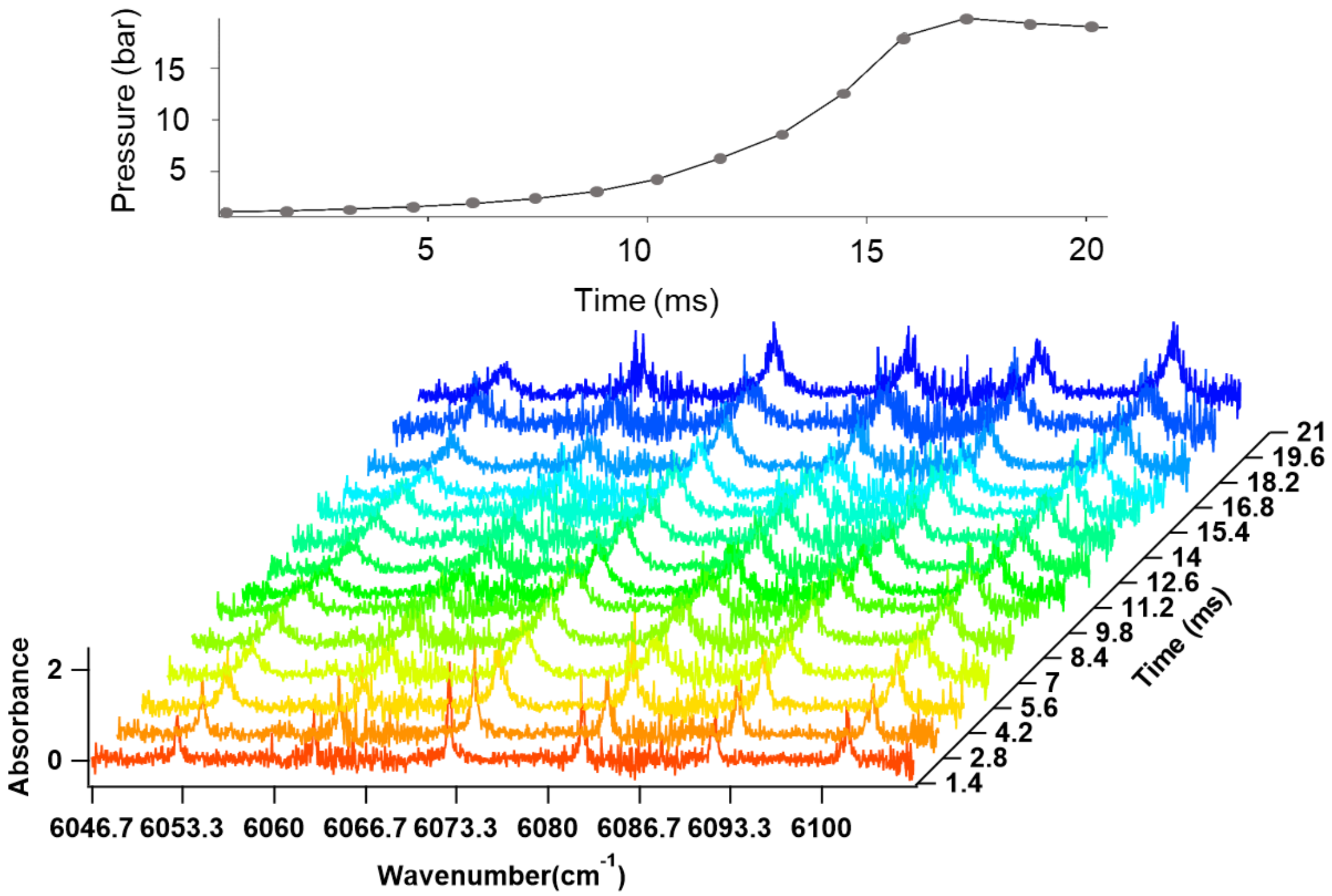

Fig 4. RCM pressure trace measured using a pressure transducer (top plot). Methane spectrum progression during RCM compression cycle (bottom plot). Each spectrum is averaged for 1.4 ms and taken at the corresponding time stamp.

## V. High speed DCS with microsecond time resolution

In order to filter both combs with control over exactly which comb teeth are being filtered, we use a cavity-comb locking scheme similar to the one in [11]. The locking scheme is based on locking the cavity to a sideband of a CW laser ($\upsilon_{cw}$). The frequency offset of the sideband from $\upsilon_{cw}$ is easily controlled by changing the drive frequency of the modulator. The CW laser is also the optical reference used for stabilizing the repetition rate of the self-referenced frequency combs. By locking one comb tooth of each comb to the reference CW laser with an offset frequency of $f_{offset}$, we create a condition in which the closest comb teeth to the CW laser map to zero Hz in the RF domain, i.e. the start of the measurement window is at $\upsilon_{cw}+f_{offset}$. Now, if we lock the cavities to the sideband of the CW reference at $\upsilon=\upsilon_{cw}+f_{offset}$, we ensure that the comb teeth corresponding to zero Hz in the RF domain are resonant with the cavities). The schematic of the experimental setup including the locking scheme is shown in Fig 5.

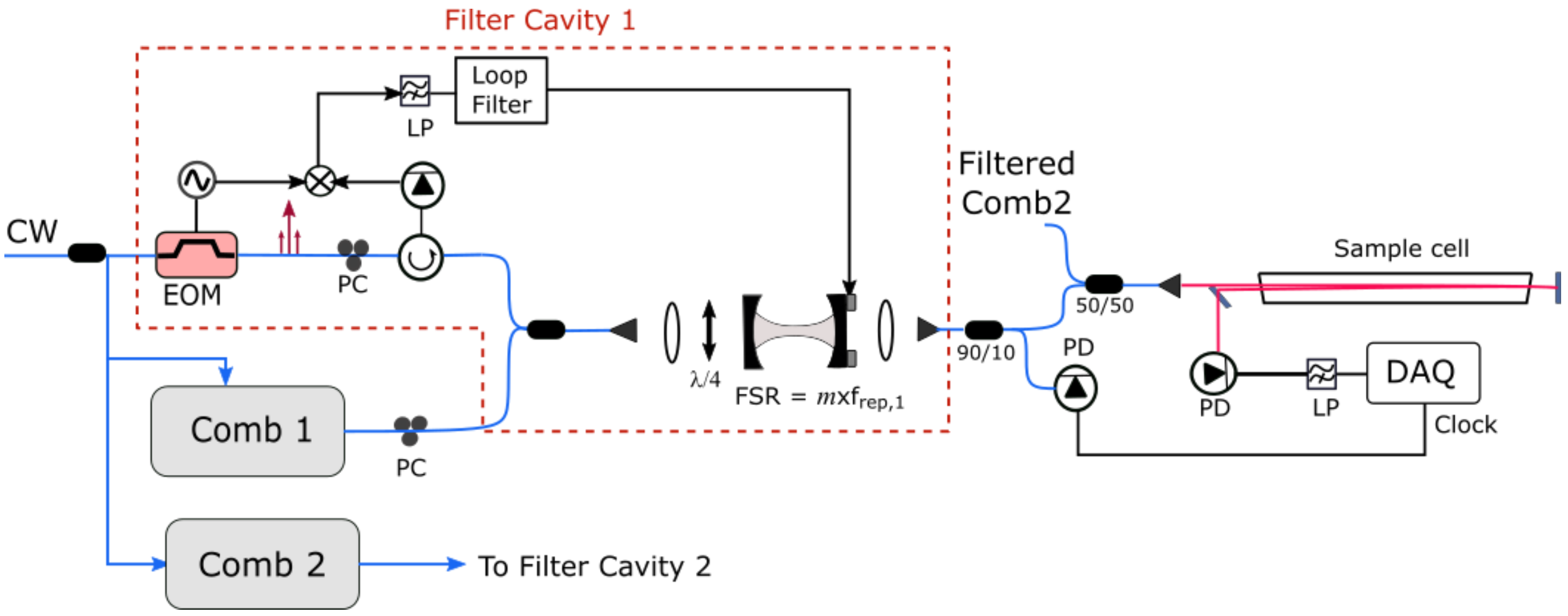

Fig.5. Schematic of the setup with two optical filters. Locking to the sideband of the CW reference laser allows for selecting the correct set of comb teeth.

In this locking scheme, some of the light from the reference CW laser is sent to the filter cavities, where it is phase modulated at $f_{offset}$, here 20 MHz, and combined with the comb light in orthogonal polarization before sending to the cavities. The Pound-Drever-Hall (PDH) error signal is obtained from the reflected light and the cavities are locked to the sideband. The combs have a repetition rate of ~204 MHz and the cavity FSR is 1.02 GHz. The filtered comb light ($f_{rep,new}$ = 1.02 GHz) is coupled into fiber. 10% of the filtered Comb 1 light is photo-detected, amplified and used as the trigger signal for the data acquisition system. The rest is combined with filtered Comb 2 light using a fiber optic 50/50 coupler. The combined light is sent to the 91-cm long spectroscopy cell containing 100% CO. After passing through the cell twice, the light is photo-detected using a 1 GHz detector and low-pass filtered to 520 MHz. The signal is then sent to a fast data acquisition card (GaGe Razormax 16) with 1 GS/s sampling rate. The difference between the repetition rate of the filtered combs is $\Delta f_{rep,new}$ ~90.9 kHz, corresponding to an acquisition time of 11 μs for each interferogram. For the filtered combs, the measurement window is 25 times larger than the native 200-MHz DCS with a $\Delta f_{rep}$ = 90.9 kHz, 183.5 $cm^{-1}$ instead of 7.3 $cm^{-1}$. Due to the imperfect overlap between the two filtered comb spectra, the high SNR portion of the recorded DCS spectrum is ~ 25 $cm^{-1}$ (750 comb teeth). Figure 6 (a) shows an example of a single shot (11 μs) absorption spectrum of CO around 6325 $cm^{-1}$ with 11.9 dB SNR obtained using the GHz DCS system.

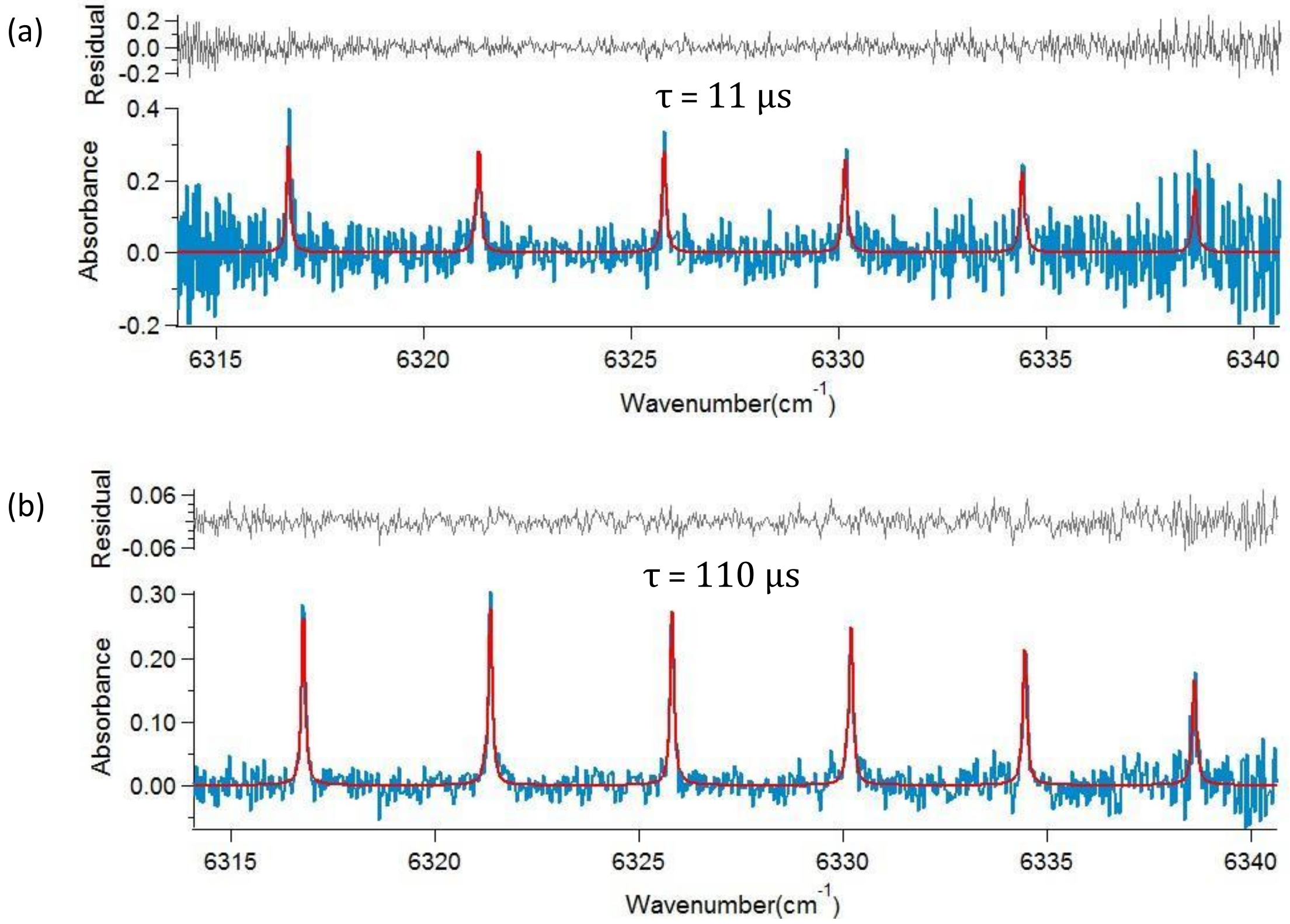

Fig. 6. Baseline corrected absorbance spectrum of CO. (a) single-shot, collected over 11 μs with 11.9 dB of SNR. (b) averaged over 110 μs with 17.1 dB of SNR.

Coherent averaging of only 10 interferograms results in even higher SNR, as seen in Fig 6 (b). The measured SNR is 17.1 dB, which is calculated from the standard deviation of the fit residual. This locking technique could be made more flexible by using two EOMs [11], one to shift the light and the other to generate phase modulation sidebands for the PDH error signal. This allows for choosing a different section of the comb spectrum to be filtered by simply changing the RF drive frequency of the EOM.

## VI. Conclusion

We demonstrated improvement of the time resolution of a 200 MHz fiber model-locked laser DCS system from seconds to milliseconds and microseconds by spectrally filtering one or both frequency combs. Spectral filtering improves the minimum acquisition time of DCS for achieving a certain SNR by allowing higher power per comb tooth and extending the photodetector linear range. Another advantage of spectral filtering is the flexibility to choose the measurement window anywhere within the octave-spanning optical bandwidth of fiber mode-locked combs. The one-filter approach has the benefit of simple implementation and is able to provide millisecond or sub-millisecond DCS time resolution. We demonstrated high-speed DCS measurements inside an RCM by spectrally filtering one of the 200-MHz frequency combs of the portable DCS system. We achieved ~3.7 times higher single-shot SNR than the unfiltered case which allowed us to collect methane spectra with 1.4-ms time resolution during a 16-ms compression cycle. The two-filter approach requires more control over filtering of the combs and as a result requires a selective locking scheme. We implemented a locking scheme that allows us to not only choose the exact comb teeth that are being filtered but also the exact optical window within the broad spectrum of the combs. Using this system, we measured a single-shot spectrum of CO encompassing 750 comb teeth around 6330 $cm^{-1}$ in 11 μs with 11.9 dB of SNR. We achieved an SNR of 17.1 dB by averaging only 10 interferograms, which corresponds to a time resolution of 110 μs. These results expand the capability of portable fiber mode-locked DCS for high-speed and high-resolution measurements in dynamic systems which require millisecond or microsecond time resolution.

**Acknowledgement:** The authors would like to thank Dr. Scott Diddams for providing the low dispersion mirrors used in fabrication of the filter cavities. We also thank Anthony Draper for assisting us with RCM experiment and Jeffrey Mohr, Andrew Zdanowicz and Prof. Anthony Marchese at Engines and Energy Conversion Laboratory at Colorado State University for operating the RCM and our funding agencies Defense Advanced Research Project Agency (W31P4Q-15-1-0011), Air Force Office of Scientific Research (FA9550-17-1-0224), National Science Foundation (CBET 1454496) and NASA Earth and Space Science Fellowship Program (18-PLANET18R).